\documentclass[a4paper,12pt]{article}
\usepackage{amsmath}
\usepackage{amsfonts}
\usepackage{amssymb}
\usepackage[usenames,x11names]{xcolor}
\usepackage{slashed}
\usepackage[pdftex]{hyperref}			% For creating hyperlinks in cross
	\hypersetup{				% references
		colorlinks = true,
		linkcolor = red,
                linktocpage = true,
		citecolor = Blue4,
		urlcolor = Green4
	}
\def\hhref#1{\href{http://arxiv.org/abs/#1}{#1}}
\def\be{\begin{equation}}
\def\ee{\end{equation}}

\def\bea{\begin{eqnarray}}
\def\eea{\end{eqnarray}}
\def\mpl{M^2}
\def\nmp{\frac{\eta}{M^2}}
\def\np2{\frac{\eta p^2}{M^2}}
\def\E1{E_e^\prime}
\def\p1{p^\prime}
\def\sw{\sin^2\theta_W}
\def\sws{\sin^4\theta_W}
\def\nk4{\frac{\eta k^4}{M^2}}
\def\ta{\tilde{\theta}}
\def\tb{\tilde{\theta_1}}
\begin{document}

\begin{titlepage}

% \rightline{UAB-FT-663}
\rightline{PRL-TH/AP-11/1}
\vskip 2.5cm
\begin{center}
{\Large {\bf  Constraint on super-luminal neutrinos from vacuum Cerenkov processes}}
\vskip 2cm
{\large Subhendra Mohanty and Soumya Rao}\\
\vskip 0.5cm
% $^1${\it Grup de F{\'\i}sica Te{\`o}rica and Institut de F{\'\i}sica
% d'Altes Energies\\
% Universitat Aut\`{o}noma de Barcelona, 08193 Bellaterra,
% Spain}\\[0.2cm]
{\it Physical Research Laboratory, Ahmedabad 380009,
India}\\
\vskip 1cm

\begin{abstract}
We examine the Cerenkov-like emission of $e^+ e^-$ from muon super-luminal muon neutrinos
assuming a quadratic energy dependence of the neutrino velocity  arising from Lorentz
violating interactions. We find that with the OPERA result for the neutrino-photon
velocity difference , the decay length for the process $\nu_\mu\rightarrow \nu_\mu e^+
e^-$ is 17,039 km which is much larger than the OPERA neutrinos path length of 730 km. We
also calculate the pion rate for super-luminal outgoing neutrinos, and we find that the
deviation of the pion decay length from the standard Lorentz conserving case at the OPERA
neutrino energy is $2\%$.
We conclude that if the muon-neutrino velocity has a quadratic energy dependence,
then OPERA result is consistent with non-observation of forbidden neutrino decays and
large deviations from the standard pion decay lifetime.
 
\end{abstract}
\end{center}
\end{titlepage}

\section{Introduction}

The OPERA experiment has recently claimed to have observed neutrinos traveling faster than
light\cite{Opera}. A possible explanation for such superluminal neutrinos comes from
Lorentz violating interactions.  An important phenomenological constraint on super-luminal
neutrinos is from the observation of Cohen and Glashow (CG) \cite{Cohen:2011hx} that
super-luminal muon neutrinos will lose energy via a Cerenkov like emission of an $e^+ e^-$
pair.

The OPERA result  find that muon neutrinos of average energy 17.5 GeV  traverse a distance
of 730 km from CERN to Gran Sasso with a velocity $c_\nu$ which exceeds the photon
velocity, \be \delta(E=17.5GeV)= \frac{(v_\nu-c)}{c}= (2.48 \pm 0.28 (stat) \pm 0.30
(syst) ) \times 10^{-5} \ee This is comparable with an earlier measurement of muon
neutrino velocity by MINOS \cite{Minos} who found that muon neutrinos of average energy 3
GeV traversing a distance 730 km exceed c by an amount, $\delta (E=3 GeV)= (5.1 \pm 2.9)
\times 10^{-5}$.  This is in contrast to the neutrino observations from supernova SN 1987a
\cite{SN1,SN2,SN3} where over a flight path of $51$ kpc, the neutrinos with energy in the
band $(7.5 -39)$ MeV all arrived within a time span of $12.4$ sec and the optical signal
arrived after $4$ hours of the neutrino signal (consistent with prediction of supernova
models) from which it is inferred that $\delta (E=15 MeV) \leq 10^{-9}$.  This implies
that the to be consistent with all observations, the neutrino velocity is energy dependent
\cite{Ellis, Giudice, Dass}.

Horava-Lifshitz theories \cite{Horava:2009uw,Visser:2009fg,Alexandre:2011kr} provide a
framework where theories are made renormalizable by the introduction  of Lorentz violating
higher derivative terms in the Lagrangian. Models of Lorentz violation which can give an
energy dependent neutrino velocity are discussed in
\cite{Mattingly:2009jf,Alexandre:2011bu}.  The Lagrangian for the Lorentz violating
neutrinos is given by
\begin{equation}
	\mathcal{L}=\overline{\Psi}\left(i\slashed{D}-m-\frac{\alpha_1}{M}(u\cdot D)^2-
	\frac{i\alpha_2}{M^2}(u\cdot D)^3(u\cdot\gamma)\right)P_L\Psi
	\label{lag}
\end{equation}
where  $u^a$ is a fixed four-vector which represents a preferred frame, thereby explicitly
breaking Lorentz invariance and the scale of Lorentz violation is determined by a large
mass $M$ and the dimensionless parameters $\alpha_1$ and
$\alpha_2$.  Starting with the Lagrangian (\ref{lag}) the dispersion relation for
neutrinos can  be derived of the form,
\begin{equation}
	E^2=p^2+m^2+\eta^\prime p^2+\frac{\eta p^4}{M^2}\label{liv}
\end{equation}
where $\eta^\prime= m \alpha_1/M$ and $\eta=2 \alpha_2$.

The neutrino velocity from eq.(\ref{liv}) is given by
\begin{equation}
	\delta=\frac{\partial E}{\partial p}-1 \simeq \frac{\eta^\prime}{2}+\frac{3\,\eta p^2}{2\,M^2}
\end{equation}
At energies $p \gg \sqrt{m M}$, we have $\eta p^2/M \gg \eta^\prime$ and the neutrino velocities increase quadratically  with energy.

 It has been pointed out, the modified dispersion relations for neutrinos make the process
 $\nu \rightarrow \nu f \bar f$ kinematically possible \cite{Cohen:2011hx,
 Mattingly:2009jf, Maccione:2011fr, Carmona} and the pion decay lifetime can get a sizable
 modification from the phase space of the outgoing super-luminal neutrino
 \cite{Cowsik:2011wv,Altschul:2011jd}.

We will assume that (a) energy and momenta are conserved in all Lorentz frames and (b) the
energy-momentum relation for neutrinos is of the form (\ref{liv}) in the lab frame. For
the electrons and other particles are assumed to be of the standard Lorentz invariant form
$E_i^2=p_i^2 +m_i^2$.

In this paper we compute the processes $\nu_\mu \rightarrow  \nu_\mu e^+ e^-$ and $\pi
\rightarrow \mu \nu_\mu$ assuming the muon-neutrino dispersion relation (\ref{liv}) in the
two  limits:
 \begin{enumerate}
  \item $E^2=p^2 (1+\eta^\prime)$ ( $\delta$ is independent of energy and we ignore
    neutrino masses) and
  \item $E^2=p^2 + \eta p^4/M^2$ ( $\delta $ is quadratic in energy).
 \end{enumerate}

Our results are as follows. Assuming $E^2=p^2 (1+\eta^\prime)$ we find that the lifetime
of the $\nu_\mu \rightarrow  \nu_\mu e^+ e^-$ is $\tau=882.9$ km for $E=17.5 GeV$ which
means that more than half of the OPERA neutrinos should decay in a 730 km flight length.
We also find that the pion in flight decay width decreases by $24\%$. These are large
effects which are not observed and this rules out non-zero $\eta^\prime$ as the source of
the super-luminality of neutrinos observed at OPERA.

On the other hand assuming $E^2= \eta p^4/M^2$ we find that the lifetime of the $\nu_\mu
\rightarrow  \nu_\mu e^+ e^-$ is $\tau=17038.6$ km for $E=17.5 GeV$ which implies that the
number of neutrinos is depleted by only 4.2\% in the course of the CERN-Gran Sasso flight.
The decrease in the pion in flight decay width is not observable.

\section{Neutrino energy loss by electron-positron pair emission}
\subsection{Energy dependent neutrino velocity}
We assume the neutrino dispersion relation $E^2=p^2+\eta p^4/M^2$ in the lab frame and calculate the decay width
 of the process $\nu(p)\to\nu(\p1) e^+(k)e^-(k^\prime)$.  The amplitude squared for this process is given by
\begin{align}
	\overline{|M|^2}=&32 G_F^2\left[(p\cdot k^\prime)(\p1\cdot k)\left(1-4\sw
	+8\sws\right)
	\right].
\label{modM}                  
\end{align}
The decay rate of the neutrino is in general given by
\begin{equation}
 \Gamma=\int\:\frac{d^3\p1}{(2\pi)^3 2E_\nu^\prime}\frac{d^3k^\prime}{(2\pi)^3
	2E_e^\prime}\frac{d^3k}{(2\pi)^3 2E_e}\frac{\overline{|M|^2}}{2E_\nu}\,
        (2\pi)^4\delta^4(p-\p1-k-k^\prime).
\label{grate}
\end{equation}
Using
\begin{equation*}
 \int\:\frac{d^3k}{2E_e}=\int\:d^4k\,\delta(k^2)\theta(k_0)
\end{equation*}
we have
\begin{equation}
 \Gamma=\frac{1}{8(2\pi)^5}\int\:\frac{d^3\p1}{E_\nu^\prime}\,
	\frac{d^3k^\prime}{E_e^\prime}\,\frac{\overline{|M|^2}}{E_\nu}\,
	\delta\left((p-\p1-k^\prime)^2\right)
\label{rate1}
\end{equation}
where we have performed the $k$ integral and imposed the $\delta$-function condition
$k=p-\p1-k^\prime$.  Without loss of generality we can choose
\begin{align*}
 p&=(E_\nu,0,0,|{\bf p}|)\\
 \p1&=(E_\nu^\prime,|{\bf \p1}|\sin\theta,0,|{\bf \p1}|\cos\theta)\\
 k^\prime&=E_e^\prime(1,\cos\phi\sin\theta_1,\sin\phi\sin\theta_1,\cos\theta_1).
\end{align*}
The argument of the $\delta$ function in eq.(\ref{rate1}) can be rewritten using the above
definitions as follows
\begin{equation}
 (p-\p1-k^\prime)^2=\left(\nmp (|{\bf p}|^3-|{\bf \p1}|^3)(|{\bf p}|-|{\bf \p1}|)-
			  |{\bf p}||{\bf \p1}|\theta^2\right)-D E_e^\prime
\end{equation}
where
\begin{equation}
 D=\nmp(|{\bf p}|^3-|{\bf \p1}|^3)+(|{\bf p}|-|{\bf \p1}|)\theta_1^2-|{\bf \p1}|\theta^2
   +2|{\bf \p1}|\theta\theta_1\cos\phi
\end{equation}
Here we have assumed that since we are dealing with high energy processes the angle of
scattering is typically very small and of the order of $\eta p^2/\mpl$, thus
dropping higher orders of $\theta$, $\theta_1$ and $\eta/\mpl$.
From here on we shall use the notation $p$ and $\p1$ to denote $|{\bf p}|$ and $|{\bf
\p1}|$, the magnitudes of the initial and final state neutrinos respectively.
Now we can rewrite the $\delta$-function in eq.(\ref{rate1}) as
\begin{equation}
 \frac{1}{D}\delta\left(E_e^\prime-\left(\nmp(p^3-{
\p1}^3)(p-\p1)-p\p1\theta^2\right)D^{-1}\right)
\label{delta}
\end{equation}

Using eqs.(\ref{delta}) in eq.(\ref{rate1}) we get
\begin{align}\nonumber
 \Gamma=&\frac{1}{512\pi^4}\int\:\p1 d\p1\,\int\:d\theta^2\,\int\:E_e^\prime dE_e^\prime\,
	\int\:d\theta_1^2\,\int\:d\phi\\&\delta\left(E_e^\prime-\left(\nmp(p^3-{\p1}^3)(p-\p1)-
        p\p1\theta^2\right)D^{-1}\right)\frac{\overline{|M|^2}}{DE_\nu}\label{rate2}
\end{align}
% Performing the $E_e^\prime$ integral to remove the delta function leaves us with
% \begin{equation}
%  \Gamma=\frac{1}{256\pi^3}\int\:dp\,\int\:d\theta^2\,\int\:d\theta_1^2\,
% 	\frac{\overline{|M|^2}}{DE_\nu}
% \end{equation}
The $\overline{|M|^2}$, using our choice of momenta and remembering that
$k=p-p^\prime-k^\prime$, becomes
\begin{align}\nonumber
 \overline{|M|^2}=&8G_F^2(\E1 E_\nu\p1)\left[(p-\E1)\theta^2\theta_1^2+\np2\left\{
 (p-\E1)\theta^2+2\E1\theta\theta_1\cos\phi+\theta_1^2\left(p\right.\right.
 \right.\\&\left.\left.\left.+\frac{{\p1}^2}{p}-2\frac{{\p1}^3}{p^2}-\E1
 +\E1\frac{{\p1}^2}{p^2}\right)\right\}\right]\left(1-4\sw +8\sws \right)\label{m2}
\end{align}

Now to fix the limits of the $\theta^2$ and $\theta_1^2$ integrals we need to find their
maximum values from the $\delta$-function condition i.e
\begin{equation}
 D\E1=\nmp(p^3-{\p1}^3)(p-\p1)-p\p1\theta^2
\end{equation}
For the maximum value of $\theta$ we set $\E1=0$ in the above equation so that
have
\begin{equation}
 \theta_{max}^2=\nmp\frac{(p^3-{\p1}^3)(p-\p1)}{p\p1}\label{tmax}
\end{equation}

And similarily setting $p^\prime=0$ and the electron energy at its maximum i.e
$E_e^\prime=p/2$ in the $\delta$-function condition we have,
\begin{equation}
 (\theta_1^2)_{max}=\np2\label{t1max}
\end{equation}

We make the following change of variables to pull out the factors of $\eta/\mpl$ and $p$
from
the integrand :
\begin{align}
 \p1\to\:x\,p,\qquad\theta^2\to\,\np2\tilde{\theta},\qquad\theta_1^2\to
 \,\np2\,\tilde{\theta_1}.
\end{align}
Using the above definitions in eq.(\ref{m2}) and substituting in eq.(\ref{rate2}) gives
the rate of electron-positron pair emission as
\begin{align}\nonumber
 \Gamma=\frac{G_F^2}{16\pi^4}\,\left(\np2 \right)^3 p^5&\left(1-4\sw +8\sws\right)
 	\int_0^1\:dx\:\int_0^\frac{(1-x)(1-x^3)}{x}\:\tilde{d\theta}&\\
	&\int_0^1\:d\tilde{\theta_1}\,\int_0^{2\pi}\,d\phi\:
	f(x,\tilde{\theta},\tilde{\theta_1},\phi)&
\end{align}
%where $f$ is complicated function of $x$, $\tilde{\theta}$ and $\tilde{\theta_1}$.
where
\begin{align*}
	f=&\frac{\left(x-(1+\ta) x^2-x^4+x^5\right)^2}{4\left(1+\tb(1-x)-\ta x-x^3
	+2\sqrt{\ta\tb}x\cos\phi\right)^4}\left[\tb(1-x)^2\left(\tb+x+\tb x
	\right.\right.\\ &\left.+2(1+\tb)x^2+2x^3+x^4\right)+\ta
	\left(\tb(1+\tb)+x-\tb^2 x-(1-\tb)x^4\right)\\ &\left.
	+2\sqrt{\ta\tb} \left(1-(1-\tb-\ta\tb)x-(1-\ta)x^3+(1-2
	\tb)x^4\right)\cos\phi\right].
\end{align*}
After numerically solving the above integral we get the following expression for the rate
of electron-positron pair emission
\begin{equation}
 \Gamma=\frac{G_F^2}{16\pi^4}\,\frac{1}{29}\, p^5\left(\np2\right)^3
	     \left(1-4\sw +8\sws\right) 
 \label{rate}
\end{equation}
The decay width written in terms of $\delta\simeq (3/2) (\eta p^2/M^2)$, is
\begin{equation}
 \Gamma=\frac{G_F^2}{54\pi^4}\,\frac{1}{29}\,p^5\,\delta^3\left(1-4\sw +8\sws\right)
\end{equation}

For OPERA neutrinos with energy $E=p=17.5 GeV$, the decay time is 
\be
\tau=\frac{1}{\Gamma}=17038.6 \,\,{\rm km/c}
\ee
which means that the neutrino number reduces to the fraction
$N/N_0=\exp(-730/17038.6)=0.958$. A $4.2 \%$ reduction in the number of muon-neutrinos in
the course of the CERN to Gran Sasso flight may be compatible with observations of the
same neutrino beam by ICARUS \cite{icarus}.

At higher neutrino energies $E_\nu \sim 500$GeV and above, a neutrino produced in a
collider will decay within the detector into hadrons and charged leptons, 
\be
\tau=0.9 \text{m} \left(\frac{500 GeV}{E_\nu}\right)^5
\ee
so it may be possible to test this dispersion relation at the LHC \cite{LHC}.

\subsection{Energy independent neutrino velocity}

 We calculate the rate for the process $\nu(p)\to\nu(\p1) e^+(k)e^-(k^\prime)$ assuming the
  dispersion relation $E^2=p^2 +\eta^\prime p^2/M^2$ which leads to a energy independent
$\delta=\eta^\prime/2$ which is the same assumption as made by Cohen and Glashow
\cite{Cohen:2011hx}.
 
And the expression for $\overline{|M|^2}$ in eq.(\ref{m2}) now becomes
\begin{align}\nonumber
 \overline{|M|^2}=&8G_F^2(\E1E_\nu\p1)\left[(p-\E1)(\eta^\prime\theta^2
 +\theta^2\theta_1^2)+2\eta^\prime\left\{(p-\p1-\E1)\theta_1^2\right.\right.\\
 &\left.\left.+\E1\theta\theta_1\cos\phi\right\}\right](1-4\sw+8\sws)
\end{align}
 
The $\delta$ function in eq.(\ref{rate1}) now becomes
\begin{equation}
 {D^\prime}^{-1}\delta\left(\E1-(\eta^\prime(p-\p1)^2-p\p1\theta^2){D^\prime}^{-1} \right)
\end{equation}
where
\be
D^\prime=\eta^\prime(p-\p1)+p\theta_1^2-\p1(\theta^2+\theta_1^2)+2\p1\theta\theta_1\cos\phi
\ee
Using the condition imposed by the $\delta$-function we once again derive the limits of
the two angular integrals.  Once more we put $\E1=0$ and $\p1=0$ in the $\delta$-function
condition to obtain maximum values of $\theta$ and $\theta_1$ respectively.
\begin{align}
 \theta_{max}^2&=\frac{\eta(p-p^\prime)^2}{pp^\prime}\\
 (\theta_1^2)_{max}&=\eta
\end{align}
And finally we change the variables of integration as before from $\p1$, $\theta$ and
$\theta_1$ to $x$, $\tilde \theta$ and $\tilde \theta_1$ respectively with
\begin{align}
 \p1\to\:x\,p,\qquad\theta^2\to\,\eta^\prime\tilde{\theta},\qquad\theta_1^2\to
 \,\eta^\prime\,\tilde{\theta_1}.	
\end{align}

In this case the rate of electron and positron emission from a neutrino decay  becomes
\begin{equation}
 \Gamma=\frac{G_F^2}{16\pi^4}\,\frac{1}{40}\,{\eta^\prime}^3 p^5 \left(1-4\sw +8\sws\right).
\end{equation}
Expressing the decay width in terms of  $\delta=\eta^\prime/2$ we obtain
\begin{equation}
 \Gamma=\frac{G_F^2}{2\pi^4}\,\frac{1}{40}\,  \delta^3 p^5 \left(1-4\sw +8\sws\right)
\end{equation}

For OPERA neutrinos with energy $E=p=17.5 GeV$, the decay time is
\be
\tau=\frac{1}{\Gamma}=882.9 \,\,{\rm km/c}
\ee
which means that the neutrino number reduces to the fraction
$N/N_0=\exp(-730/882.9)=0.437$. A $56\%$ reduction in the number of muon-neutrinos in the
CERN to Gran Sasso flight can safely be ruled out\cite{icarus}. This implies that the
dispersion relation $E^2=p^2(1+\eta^\prime)$ to describe energy independent super-luminal
neutrino velocities can be ruled out as pointed out in \cite{Cohen:2011hx}.

\section{ Pion decay lifetime }

\subsection{Energy dependent neutrino velocity}
We calculate the pion decay width in the lab frame  with a super-luminal neutrino in the
final state. We assume the dispersion relation $E^2= (p^2+\eta p^4/M^2)$ in the lab
frame.   The amplitude squared for the process $\pi^-(q)\to\mu^-(p)\bar{\nu}_\mu(k)$ is,
\begin{equation}
 \overline{|M|^2}=2G_F^2f_\pi^2m_\mu^2\left[m_\pi^2-m_\mu^2+\nk4
 \left(\frac{m_\pi^2}{m_\mu^2}+2\right)\right]
\end{equation}

The decay width is then given by
\begin{align}\nonumber
 \Gamma=&\frac{G_F^2f_\pi^2m_\mu^2}{16\pi^2 E_\pi}\int\,\frac{d^3p}{E_\mu}\frac{d^3 k}{k}
        \delta^3(\vec{q}-\vec{p}-\vec{k})\delta(E_\pi-E_\mu-E_\nu)\left[m_\pi^2-m_\mu^2
        \right.\\ &\left.
        +\nk4\left(\frac{m_\pi^2}{m_\mu^2}+2\right)\right]
\end{align}

Performing the $d^3p$ integral to remove the 3-momentum $\delta$-function and writing
$E_\mu=\sqrt{|\vec{q}-\vec{k}|^2+m_\mu^2}$ the decay rate then becomes
\begin{align}
 \Gamma=&\frac{G_F^2f_\pi^2m_\mu^2}{8\pi E_\pi}\int\,
	\frac{k\,dk\,d\cos\theta}{\sqrt{|\vec{q}-\vec{p}|^2+m_\mu^2}}
	\delta(E_\nu+\sqrt{|\vec{q}-\vec{k}|^2+m_\mu^2}-E_\pi)\nonumber\\
	&\left[m_\pi^2-m_\mu^2+\nk4\left(\frac{m_\pi^2}{m_\mu^2}+2\right)\right]
\label{pi-rate}
\end{align}

Writing $|\vec{q}-\vec{k}|^2=k^2+q^2-2kq\cos\theta$, $\theta$ being the angle between
$\vec{k}$ and $\vec{q}$, and $E_\nu=k+\eta k^3/(2M^2)$ we see from the argument of the
$\delta$-function in eq.(\ref{pi-rate})
\begin{equation}
%	\cos\theta=\frac{m_\mu^2-m_\pi^2+2E_\pi k+\frac{\eta k^3}{M^2}E_\pi-\frac{\eta
%	k^4}{M^2}}{2kq}\label{cos}
\cos\theta=\left(m_\mu^2-m_\pi^2+2E_\pi k+\frac{\eta k^3}{M^2}E_\pi-\frac{\eta
k^4}{M^2}\right)(2kq)^{-1}\label{cos}
\end{equation}
while the derivative of the argument of $\delta$-function with respect to $\cos\theta$
yields
\begin{equation}
\left|\frac{d}{d\cos\theta}(E_\nu+\sqrt{|\vec{q}-\vec{k}|^2+m_\mu^2}-E_\pi)\right|=
\frac{kq}{\sqrt{|\vec{q}-\vec{k}|^2+m_\mu^2}}
\end{equation}
Substituting this in eq.(\ref{pi-rate}) we get
\begin{equation}
 \Gamma=\frac{G_F^2f_\pi^2m_\mu^2}{8\pi E_\pi}\int\,\frac{dk}{q}
        \left[m_\pi^2-m_\mu^2+\nk4\left(\frac{m_\pi^2}{m_\mu^2}+2\right)\right]
\end{equation}

The limits of the $k$ integral are fixed by taking $\cos\theta=\pm 1$ in eq.(\ref{cos})
\begin{align}
	k_{max}=&\frac{m_\pi^2-m_\mu^2-\dfrac{\eta k_{max}^3}{M^2}(E_\pi-k_{max})}{2(E_\pi-q)}\\%[8pt]
	k_{min}=&\frac{m_\pi^2-m_\mu^2-\dfrac{\eta k_{max}^3}{M^2}(E_\pi-k_{min})}{2(E_\pi+q)}
% \label{kmax}
\end{align}
we solve these polynomial equations for $k_{max}$ and $k_{min}$ numerically to obtain the
kinematically allowed limits of neutrino momentum.  Using these limits to integrate over
the neutrino momentum $k$ we get the decay rate for pion.  The effect of the superluminal
neutrinos here is to restrict the phase space by restricting $k_{max}$ and $k_{min}$.  As a
result the ratio of the pion decay width to the Standard Model prediction
\begin{equation}
	\Gamma_0(\pi\to \mu \nu)=\frac{m_\pi^2}{E_\pi}\left(1-\frac{m_\mu^2}{m_\pi^2}\right)^2
	\label{}
\end{equation}
is found to be $\dfrac{\Gamma}{\Gamma_0}=0.98$ for $E_\pi= 20$ GeV.	

However for a $100$ GeV pion the reduction can be as large as $73\%$.

%\begin{equation}
% \Gamma=\frac{G_F^2f_\pi^2m_\mu^2}{8\pi}\left[\frac{m_\pi^2}{E_\pi}\left(1-\frac{m_\mu^2}
% {m_\pi^2}\right)^2+\frac{\eta E_\pi^3}{5\mpl}\left(1-\frac{m_\mu^2}{m_\pi^2}\right)^5
% \left(\frac{m_\pi^2}{m_\mu^2}+2\right)\right]
%\end{equation}
%The additional contribution from the Lorentz violating terms to the standard pion decay
%width $\Gamma_0$ can be written as
%\bea
% \frac{\Gamma-\Gamma_0}{\Gamma_0}&=&\frac{1}{5}\left(\frac{\eta E_\pi^2}{\mpl}\right)
% \left(\frac{E_\pi^2}{m_\pi^2}\right)\left(1-\frac{m_\mu^2}
%{m_\pi^2}\right)^3\left(\frac{m_\pi^2}{m_\mu^2}+2\right)\nonumber\\ 
%&=& 0.02 \left(\frac{E_\pi}{20 GeV} \right)^2
%\eea
%Hence the correction to the pion decay width is only $2\%$.

\subsection{Energy independent neutrino velocity} 
  We  now assume the dispersion relation $E^2= (1+\eta^\prime)p^2$ in the lab frame and
calculate the pion decay width.  The amplitude squared for the process
$\pi^-(q)\to\mu^-(p)\bar{\nu}_\mu(k)$ is
\begin{equation}
 \overline{|M|^2}=2G_F^2f_\pi^2m_\mu^2\left[m_\pi^2-m_\mu^2
                  +\eta^\prime k^2\left(\frac{m_\pi^2}{m_\mu^2}+2\right)\right]
\end{equation}

The decay width is then given by
\begin{align}\nonumber
 \Gamma=&\frac{G_F^2f_\pi^2m_\mu^2}{16\pi^2 E_\pi}\int\,\frac{d^3p}{E_\mu}\frac{d^3 k}{k}
        \delta^3(\vec{q}-\vec{p}-\vec{k})\delta(E_\pi-E_\mu-E_\nu)\left[m_\pi^2-m_\mu^2
        \right.\\ &\left.
        +\eta^\prime k^2\left(\frac{m_\pi^2}{m_\mu^2}+2\right)\right]
\end{align}

Using the same procedure as in the last section we find the limits for the $k$ integral to
be
\begin{align}
	k_{max}=&\frac{m_\pi^2-m_\mu^2-\eta^\prime k_{max}(E_\pi-k_{max})}{2(E_\pi-q)}\\[8pt]
 	k_{min}=&\frac{m_\pi^2-m_\mu^2-\eta^\prime k_{min}(E_\pi-k_{min})}{2(E_\pi+q)}
% \label{kmax}
\end{align}
Solving these equations gives the following expressions for $k_{max}$ and $k_{min}$
\begin{align}
	k_{max}={\eta^\prime}^{-1}\left(E_\pi-q+\frac{\eta E_\pi}{2}-\Delta_-\right)\\
	k_{min}={\eta^\prime}^{-1}\left(E_\pi+q+\frac{\eta E_\pi}{2}-\Delta_+\right)
	\label{}
\end{align}
where
\begin{align}
	\Delta_+=\sqrt{\left(E_\pi+q+\frac{\eta^\prime E_\pi}{2}\right)^2-\eta^\prime
	(m_\pi^2-m_\mu^2)}\\
	\Delta_-=\sqrt{\left(E_\pi-q+\frac{\eta^\prime E_\pi}{2}\right)^2-\eta^\prime
	(m_\pi^2-m_\mu^2)}
	\label{}
\end{align}
Integrating over these limits gives the decay width for pion in flight as
\begin{align}\nonumber
	\Gamma=\frac{G_f^2 f_\pi^2 m_\mu^2}{8\pi q E_\pi}&\left[{\eta^\prime}^{-1}\left(
	m_\pi^2-m_\mu^2\right)\left(\Delta_+-\Delta_--2q\right)\right.\\&\left.
	+\eta^\prime\left(\frac{m_\pi^2}{m_\mu^2}+2\right)\left(k_{max}^3-k_{min}^3\right)\right]
	\label{}
\end{align}

In this case the reduction in pion decay width compared to the Standard Model prediction
for an incident pion of energy $20$ GeV is found to be $32\%$ while for $100$ GeV incident
energy it can be as large as $96\%$.

%\begin{equation}
% 	\Gamma=\frac{G_F^2f_\pi^2m_\mu^2}{8\pi}\left[\frac{m_\pi^2}{E_\pi}\left(1-\frac{m_\mu^2}
% 	{m_\pi^2}\right)^2+\frac{\eta^\prime E_\pi}{3}\left(1-\frac{m_\mu^2}{m_\pi^2}\right)^3
% 	\left(\frac{m_\pi^2}{m_\mu^2}+2\right)\right]
%	\label{}
%\end{equation}

%The additional contribution from the Lorentz violating terms to the standard pion decay width $\Gamma_0$ can be written as
%\bea
% \frac{\Gamma-\Gamma_0}{\Gamma_0}&=&\frac{\eta^\prime}{3}\left(\frac{ E_\pi^2}{m_\pi^2}\right)
% \left(1-\frac{m_\mu^2}{m_\pi^2}\right)\left(\frac{m_\pi^2}{m_\mu^2}+2\right)\nonumber\\
%&=& 0.543 \left(\frac{E_\pi}{20 GeV} \right)^2
%\eea
%This is a large correction consistent with the estimate in \cite{Cowsik:2011wv} which
%rules out the dispersion relation $E=(1+\eta^\prime/2) p$ for describing the OPERA
%observations.

\section{Conclusions}

We calculate the decay width forbidden process $\nu\to \nu e^- e^+$ which is allowed if
the neutrino has a dispersion relation $E^2=m^2 +p^2 +\eta^\prime p^2 + (\eta/M^2) p^4$ in
the context of the superluminal neutrinos observed at OPERA. We find that when the
dispersion relation is dominated by the $\eta$ term (and the neutrino velocity $(v_\nu -1)
\propto  E_\nu^2$) then the mean decay length for this process is larger than the OPERA
neutrino flight path.  When the path length of the neutrinos is much larger than the
calculated decay length then we can use the relation $dE/dx=\Gamma E$ to calculate the
energy loss rate.  This relation is only valid if there are multiple decays from a single
neutrino over the path length.  However for Opera neutrino energies we find that for the
$\eta$ dominant dispersion relation the decay length is 17000 km(much larger than the
Opera path length).  In this case the energy of the neutrino beam  $<E(L)>= E_0
\exp(-L/\tau)$, which results in only 4 percent reduction in energy.  We also compute the
pion decay width for these outgoing neutrinos and find that the deviation from the
standard Lorentz conserving case is $2 \%$.

We also compute these processes assuming that the $\eta^\prime$ term in the dispersion
relation dominates (and $v_\nu-1$ is independent of neutrino energy) and find that the
decay length for the $\nu\rightarrow \nu e^+ e^-$is smaller than the OPERA neutrino path
and this possibility can be ruled out by Cohen and Glashow \cite{Cohen:2011hx}. We also
find that the pion decay width is reduced by $32 \%$ and this possibility for the
dispersion relations can be ruled out as  pointed out in
\cite{Cowsik:2011wv,Bi:2011nd}.  We have worked in framework of explicit
Lorentz violation in the Lagrangian which gives a frame dependent dispersion relation.
There exists other possibilities for generalising Lorentz transformation such that the
modified dispersion relations are also covariant and these theories too can evade the
constraints of Cerenkov processes \cite{AmelinoCamelia:2011bz}.  A numerical calculation
of these processes using generalised dispersion relations has been performed in
\cite{Bi:2011nd}.  However in \cite{Bi:2011nd} the calculation was done in the centre of
mass frame of the outgoing particles.  But this calculation should be performed in the lab
frame where eq.(\ref{liv}) is valid since the dispersion relation is not frame
independent.

We conclude that the OPERA neutrino measurement is only compatible with the energy
dependent neutrino velocity, and this possibility can be tested at the LHC with neutrinos
produced with energies above 500 GeV.

\end{document}